\documentclass{article}

\usepackage[final]{neurips_2022}

\usepackage[utf8]{inputenc} 
\usepackage[T1]{fontenc}    
\usepackage{hyperref}       
\usepackage{url}            
\usepackage{booktabs}       
\usepackage{amsfonts}       
\usepackage{nicefrac}       
\usepackage{microtype}      
\usepackage{xcolor}         
\usepackage{graphicx}
\usepackage{float}
\usepackage{wrapfig}

\usepackage{natbib}
\DeclareUnicodeCharacter{2212}{-}

\usepackage{amsmath}

\usepackage{caption}
\usepackage{subcaption}

\title{Influencer Detection with Dynamic Graph Neural Networks}

\author{%
  Elena Tiukhova \\
  Research Center for Information Systems \\ Engineering \\
  Faculty of Economics and Business \\
  KU Leuven\\
  Naamsestraat 69, 3000 Leuven, Belgium \\
  \texttt{elena.tiukhova@kuleuven.be} \\
 \And
 Emiliano Penaloza \\
 Department of Statistical and Actuarial Sciences\\
 The University of Western Ontario \\
 1151 Richmond Street, \\ London, Ontario N6A 5B7, Canada \\
 \texttt{epenaloz@uwo.ca} \\
 \And
 Hernan Garcia \\
 Rappi \\
Cl.93 \#19-58, Bogotá, Colombia \\
 \texttt{javier.garcia@rappi.com} \\
 \And
 Alejandro Correa Bahnsen \\
 Rappi \\
Cl.93 \#19-58, Bogotá, Colombia \\
 \texttt{alejandro.correa@rappi.com} \\
 \And
  María Óskarsdóttir \\
 Department of Computer Science \\
 Reykjavík University \\
 Menntavegi 1, 102 Reykjavík, Iceland  \\
 \texttt{mariaoskars@ru.is} \\
  \And
 Bart Baesens \\
 Research Center for Information Systems \\ Engineering \\ Faculty of Economics and Business \\
  KU Leuven\\
  Naamsestraat 69, 3000 Leuven, Belgium \\
 \texttt{bart.baesens@kuleuven.be} \\
  \And
 Monique Snoeck \\
 Research Center for Information Systems \\ Engineering \\ Faculty of Economics and Business \\
  KU Leuven\\
  Naamsestraat 69, 3000 Leuven, Belgium \\
 \texttt{monique.snoeck@kuleuven.be} \\
 \And
 Cristián Bravo \\
 Department of Statistical and Actuarial Sciences\\
 The University of Western Ontario \\
 1151 Richmond Street \\ London, Ontario N6A 5B7, Canada \\
 \texttt{cbravoro@uwo.ca} \\
}

\begin{document}

\maketitle

\begin{abstract}
Leveraging network information for prediction tasks has become a common practice in many domains. Being an important part of targeted marketing, influencer detection can potentially benefit from incorporating dynamic network representation. In this work, we investigate different dynamic Graph Neural Networks (GNNs) configurations for influencer detection and evaluate their prediction performance using a unique corporate data set. We show that using deep multi-head attention in GNN and encoding temporal attributes significantly improves performance. Furthermore, our empirical evaluation illustrates that capturing neighborhood representation is more beneficial that using network centrality measures.
\end{abstract}

\section{Introduction}
\label{introduction}

Advances in data collection and processing have enhanced the use of automated data workflows for decision-making. A primary source of data comes in the form of networks that capture connections between people. When relational information is leveraged, it is assumed that people in the network influence each other's behavior and decisions, which has been shown to be true in many domains such as fraud detection \citep{baesens2015fraud} or e-commerce recommendations \citep{SUN2015109}.

A common way information flows through a network is by the  Word-of-Mouth effect which is seen as a powerful tool for spreading influence among customers in marketing \citep{puigbo2014influencer}. Customers who succeed in utilizing this effect in order to change others perspectives are considered influencers \citep{rogers1962methods}. It is possible to model such scenarios as a network, in which its topology plays a crucial role in identifying influencers. This is a large area of study with many standard approaches for encoding the network topology, such as neighborhood and centrality metrics as well as collective inference algorithms \citep{baesens2015fraud}. Due to the rising popularity of deep learning, graph neural networks (GNNs) are extensively used for end-to-end tasks of graph learning \citep{rhee2017hybrid, guo2019attention}. However, the research on influencer detection with GNNs is limited, especially when it comes to networks that evolve in time and when there exist several types of connections in a network. To the best of our knowledge, there exists no research about influencer detection on dynamic attributed edge-colored networks with GNNs. 

The main purpose of this paper is to add to the body of research on influencer detection by evaluating different dynamic GNNs configurations and to investigate whether encoding network topology using GNNs together with capturing its dynamic evolution have an added value for performance. By doing so, the following contributions are made. Firstly, we adapt dynamic GNNs for ex-post influencer detection, that is, identifying current users of the product or service who influence neighboring non-users to acquire it in the future. Secondly, we evaluate different GNN configurations in combination with different RNN configurations for our problem\footnote{The code is available at \url{https://github.com/Banking-Analytics-Lab/DynamicGraphLearning}}. Finally, we compare the results to baseline static graph neural networks and dynamic non-GNN approaches.

\section{Related work}
\cite{puigbo2014influencer} highlight the importance of influencer detection since the rise of the Internet. However, most of the traditional approaches lack more advanced indicators of the relationships between network actors. Due to the rising popularity of GNNs and the demand for improved relationship extraction techniques, the graph influence network framework has been proposed by \cite{shi2022graph}. The framework is aimed at finding the influential neighbors of a node. However, it is not designed for detecting global influencers and can be applied on static networks only. 

Networks can be seen as an unstructured data source, requiring specific methods in order to extract network topology and be able to incorporate it into prediction models. Some approaches to learning on networks are based on matrix factorization including spectral clustering \citep{von2007tutorial} and learning with modularity matrix \citep{tang2009relational}. More advanced methods are based on learning by performing random walks on a networks, e.g., DeepWalk \citep{perozzi2014deepwalk} and node2vec \citep{grover2016node2vec}. Network topology can be incorporated into the model by extracting centrality information using PageRank-like algorithms which have been proved to be beneficial for, e.g., credit risk prediction in multilayer networks \citep{oskarsdottir2021multilayer}. The enhancements in deep learning have brought GNNs to the forefront of the field where they demonstrate cutting-edge performance \citep{zhou2020graph}. The general design pipeline of GNNs includes the steps of specifying the network type and scale, deciding on the task type and building the model using carefully designed computational modules \citep{zhou2020graph}.

\section{Influencer detection with Discrete Time Dynamic Graphs}
Following the design pipeline of \cite{zhou2020graph}, we define the task of future influencer detection in this paper as a supervised node-level learning problem on a dynamic heterogeneous undirected network. A typical network learning process consists of an encoder and decoder  \citep{hamilton2017representation}. The encoder part of the model is aimed at learning node embeddings while the decoder part is used to solve a prediction task, e.g., node classification. \citet{survey} survey different encoder-decoder architectures that exist for supervised dynamic graph learning and classify them into Discrete Time Dynamic Graph (DTDG) learning that uses network snapshots and Continuous Time Dynamic Graph (CTDG) learning that deals with an updating event stream in a network, e.g., the Temporal Graph Networks framework for deep learning on dynamic network represented as sequences of timed events \citep{rossi2020temporal}. Following their taxonomy, we capture the network topology at each timestamp by applying the DTDG encoder for attributed static networks, namely, Graph Convolutional Networks (GCNs) \citep{kipf2017semisupervised} and Graph Attention Networks (GATs) \citep{velichkovic2018graph}. We capture the dynamic nature of the networks by employing the models from the RNN family as a DTDG decoder. We also compare these models with baseline models, namely, the PageRank+RNN, static GNNs and dynamic non-GNN models. 

\subsection{GNN + RNN}
\label{GNNs}
In order to deal with complex data structures and arbitrary size of neighborhood, \cite{kipf2017semisupervised} propose Graph Convolutional Networks that can learn on non-euclidean data such as networks. A GCN model learns node embeddings by aggregating information from a node's neighborhood. However, it assigns the same importance to all the neighboring nodes which is rarely the case in practice. Hence, \cite{velichkovic2018graph} introduce Graph Attention Networks where nodes follow a self-attention mechanism and assign an importance to each connection by attending over their neighbors.

The disadvantage of both GCN and GAT architectures is that they are static and do not take into account the dynamic changes in the network that happen over time. To take into account the time dimension, we use the GNN models mentioned above together with Recurrent Neural Networks, namely the Long Short-Term Memory (LSTM) model \citep{hochreiter1997long} and Gated Recurrent Units (GRUs) model \citep{cho2014learning}. The LSTM model is capable of learning long-term dependencies by storing long-term memory in a cell state while capturing the most recent information in its hidden states. The GRU model is a less complex version of LSTMs as it does not have a cell state and stores long-term memory directly in its hidden states. In both the LSTM and GRU models, the output embeddings of a GNN model are used as an input.

\begin{figure}[H]
     \centering
     \begin{subfigure}[b]{0.45\textwidth}
          \centering
  \includegraphics[width=0.8\linewidth]{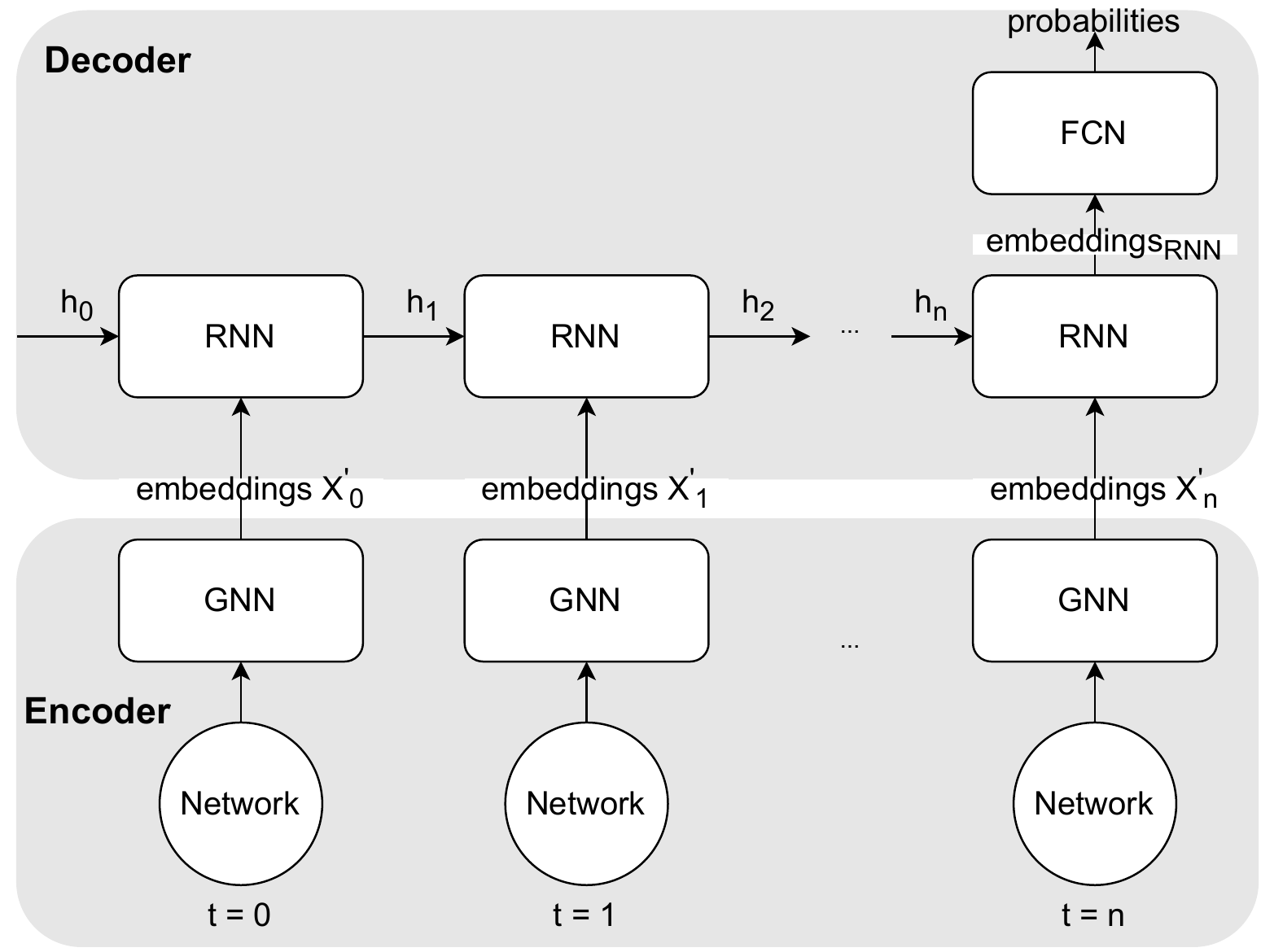}
  \caption{GNN+RNN}
  \label{fig:GNN}
     \end{subfigure}
     \hfill
     \begin{subfigure}[b]{0.45\textwidth}
         \centering
  \includegraphics[width=0.8\linewidth]{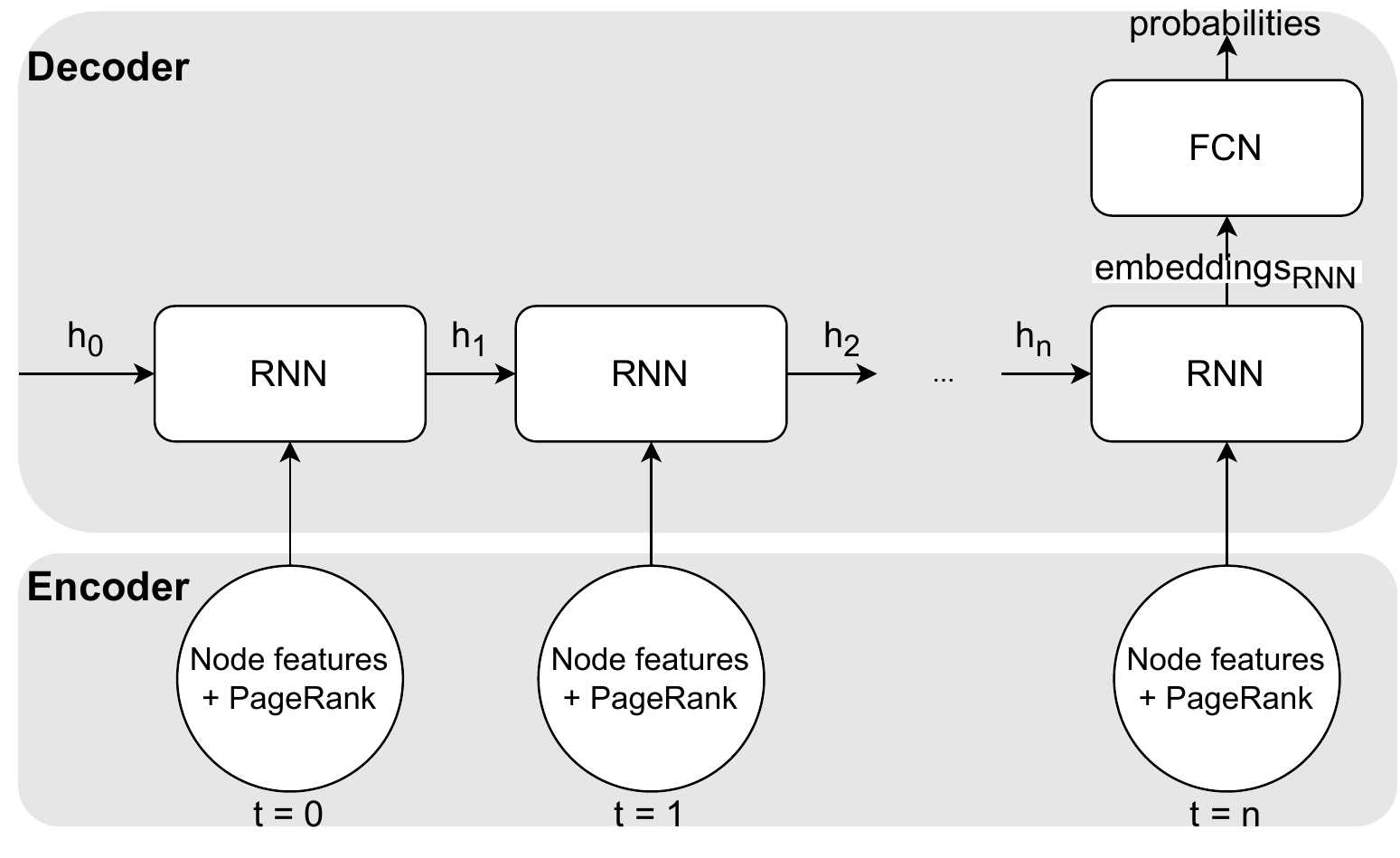}
  \caption{Features(+PageRank)+RNN}
  \label{fig:PageRank}
     \end{subfigure}
     \caption{Models' architecture}
     \label{fig:architectures}
\end{figure}

Considering all the above, we investigate four different model configurations, i.e., GCN+LSTM, GCN+GRU, GAT+LSTM, and GAT+GRU. Their architecture is displayed in Figure \ref{fig:architectures}.

\subsection{PageRank + RNN \& Features + RNN}
Following the strategy used in the research by \cite{oskarsdottir2017social, oskarsdottir2022social}, we can enrich non-relational 
classifiers with network features using feature engineering \citep{verdonck2021special}. One of such network features that can summarize node importance is PageRank \citep{brin1998anatomy}. As the PageRank value represents the relative importance of the node within one component, we calculate the value for PageRank separately within each of the connected components. PageRank values are used as an additional node feature and subsequently utilized for dynamic node classification with LSTMs and GRUs (Figure \ref{fig:PageRank}). 

\section{Experiments}
\label{experiments}
\begin{wrapfigure}{r}{0.5\textwidth}
\centering
  \centering
  \includegraphics[width=0.48\textwidth]{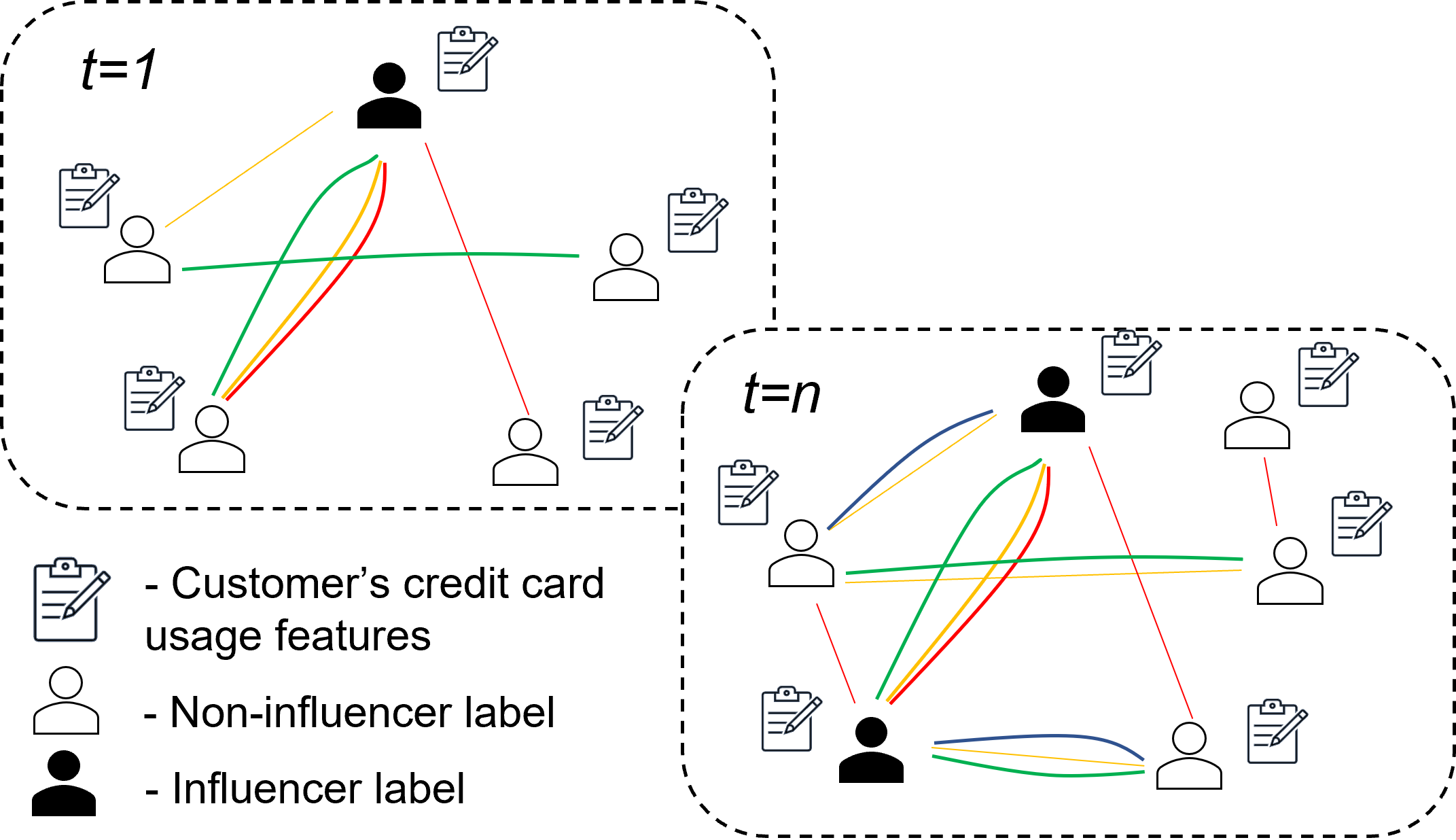}
  \caption{Network}
  \label{fig:Network}
\end{wrapfigure}

We utilize data from one city of a Super-App company operating in Latin America that has both a delivery app and issues credit cards to its customers. We construct monthly snapshots based on different types of connections between the users (Figure \ref{fig:Network}), resulting in a dynamic attributed undirected network with implicit time and colored edges. The nodes in the network represent the customers and are attributed with the features that characterize the customer's credit card usage in each monthly snapshot. The edges in the network are colored by different types of connections between network actors as displayed in Appendix \ref{sec:appendices}, Table \ref{Edge_types}. Coloring the edges is performed by creating edge features (transformed to edge weights in the GCN+RNN models): three binary edge features for credit card, geohash and contacts edge types with the last one being enriched with the references edges created in the network snapshot the month following the month of referral. We note that existing connections never disappear from the network while it is possible that new connections are established (Appendix \ref{sec:appendices}, Figure  \ref{fig:Network_characteristics}). The network is labelled: customers who referred (i.e., extended an invitation to the Super-App's services to someone they are connected to) other customers in the past at least once are labelled as influencers while the remaining customers are labelled as non-influencers (see Figure \ref{fig:Network}).

The network is imbalanced, as being an influencer is less common than being a non-influencer (imbalance ratio is ${\sim}13\%$). We check if oversampling helps to handle the data imbalance during the model validation step and apply the Synthetic Minority Oversampling TEchnique (SMOTE) \citep{chawla2002smote} to the embeddings generated by the RNN model. We follow the oversampling strategy of \cite{zhao2021graphsmote} and oversample the nodes in the embeddings generated by the RNN model or by the GNN encoder in static GNN models. We generate synthetic nodes of the minority class of different quantities that we set as a hyperparameter (Appendix \ref{sec:appendices}, Table \ref{Hyperparameters}). 

The data splits, a general pipeline of training, validating and testing the models as well as the best hyperparameter specifications found by grid search are displayed in Appendix \ref{sec:appendices}, Figure~\ref{fig:Model_implementation} and Table \ref{Hyperparameters}, respectively. The type of resources used is displayed in Appendix \ref{sec:appendices}, Table \ref{resources}.
  
\section{Results and Discussion}
\label{results}

\begin{table}[H]
\fontsize{8}{10}\selectfont 
  \caption{Models' performance (models are implemented using the pytorch geometric library \citep{Fey/Lenssen/2019}). Confidence intervals of AUC values are obtained from bootstrapping AUC values.}
  \label{Models_performance2}
  \centering
  \begin{tabular}{p{2cm}p{2.6cm}p{2.6cm}p{2.6cm}}
    \toprule                 
    Model & Test AUC seen nodes & Test AUC unseen nodes & Total time (seconds) \\
    \midrule
    GCN+LSTM & 0.756$\pm$0.006 & 0.786$\pm$0.014 & 12747.2 \\
    GCN+GRU & 0.843$\pm$0.005 &  0.730$\pm$0.015 & 12752.5 \\
    GAT+LSTM & 0.842$\pm$0.005 & \textbf{0.831$\pm$0.012} &  58919.6  \\
    GAT+GRU & \textbf{0.864$\pm$0.004} & \textbf{0.823$\pm$0.013} &  58801.3  \\
    PageRank+LSTM & 0.672$\pm$0.007 & 0.685$\pm$0.009 & 2844.3 \\
    PageRank+GRU & 0.801$\pm$0.004 & 0.665$\pm$0.015 & 4637.8 \\
   Features+LSTM & 0.673$\pm$0.006 & 0.686$\pm$0.009 & 2275.4 \\
   Features+GRU & 0.799$\pm$0.006 & 0.673$\pm$0.009 & 4635.3 \\
   Static GAT & 0.639$\pm$0.005 & 0.700$\pm$0.016 & 40142.5	 \\
   Static GCN  & 0.635$\pm$0.006 & 0.663$\pm$0.024 & 1742.3 \\
    \bottomrule
  \end{tabular}
\end{table}

As can be seen from Table \ref{Models_performance2}, GNN+RNN models in general outperform baseline models with the GAT+GRU configuration being the best one on both seen and unseen nodes and the GAT+LSTM being statistically identical to GAT+GRU over unseen nodes. A notable improvement over baseline models is obtained on unseen nodes with an AUC increase of 0.13 obtained on the best GAT+LSTM model. We also note that the models with GAT as an encoder outperform the models with a network topology encoded by GCN.

The best GCN+RNN configurations consist of 200 embeddings generated by GCN with one hidden layer and 200 hidden dimensions in both GCN and RNN (LSTM or GRU). In contrast, the best GAT+RNN configurations are deep GATs with 4 layers and 4 heads generating 200 embeddings and 100 hidden layers in GAT and RNN (Appendix \ref{sec:appendices}, Table \ref{Hyperparameters}). Also, we note that upsampling does not increase the performance meaning that most of the models can deal with the data imbalance. Therefore, using deep multi-head attention mechanism helps to better capture network topology than just aggregating information from a node’s neighborhood over both balanced and unbalanced sets.

Among the baseline models, non-GNN dynamic models consistently outperform non-dynamic GNNs on seen nodes while the static GAT model being the best over unseen nodes. Hence, capturing time-evolving patterns plays an important role in predicting future influencers among seen nodes while the network typology encoding is crucial for generalizing to unseen nodes. Moreover, adding PageRank as an additional feature does not result in a significant performance improvement. Thus, neighbor feature representations captured by GNNs are more important for influencer detection than using centrality measures such as PageRank.

\section{Conclusion}
\label{conclusion}

Early detection and targeting of influencers allows for efficient spread of information through the network. Hence, different model architectures for influencer detection with networks should be evaluated. For these reasons, we researched different dynamic GNNs configurations and investigated whether encoding network topology with GNNs and capturing the dynamic evolution of the network have an added value to the prediction performance. 

First, neighbor feature representations captured by GNNs are more important than centrality measures such as PageRank especially when it comes to generalizing to unseen nodes where using multi-head attention in the encoder boosts the performance. Second, we conclude that dynamics of the network plays an important role; thus, the decoder of the model should capture time. As the use of the influencer detection model is intended for marketing, we foresee the best models will allow optimizing the frequency and tenor of targeted marketing actions some users can be subjected to.

Our work has a few limitations. First, there could exist other connections in the network that are not captured by a current network setup. Moreover, the way edges are created based on the Geohash 7 (Appendix \ref{sec:appendices}, Table \ref{Edge_types}) proximity can affect the connectivity strength of the network. Future improvements include unsupervised learning algorithms for influencer detection including anomaly detection methods for dynamic networks. Next, we can evaluate models from a profit-driven perspective that will enable us to bring more business context into the results. Furthermore, we can explore behavioral node features from the delivery app which can potentially increase predictive power. Moreover, we can expand the network to more than one city to study how generalizable the results are in a larger geographical area. The future research avenues also include exploring more encoder configurations such as GraphSAGE \citep{hamilton2017inductive} or Graph Isomorphism Networks \citep{xu2018powerful}.  

\section{Acknowledgements}
The research was sponsored by the ING Chair on Applying Deep Learning on Metadata as a Competitive Accelerator. The fifth author acknowledges the support of the Icelandic Research Fund (IRF) [grant number 228511-051]. The last author acknowledges the support of the Natural Sciences and Engineering Research Council of Canada (NSERC) [Discovery Grant RGPIN-2020-07114]. This research was undertaken, in part, thanks to funding from the Canada Research Chairs program. This research was enabled in part by support provided by Compute Ontario (\url{computeontario.ca}) and the Digital Research Alliance of Canada (\url{alliancecan.ca}) [FT \#2070].

\bibliography{neurips_2022}

\medskip

\appendix
\renewcommand\thefigure{\thesection.\arabic{figure}}
\renewcommand\thetable{\thesection.\arabic{table}}    
\setcounter{table}{0}    
\setcounter{figure}{0}    

\section{Appendix}\label{sec:appendices}

\begin{figure}[H]
\centering
  \centering
  \includegraphics[width=0.9\linewidth]{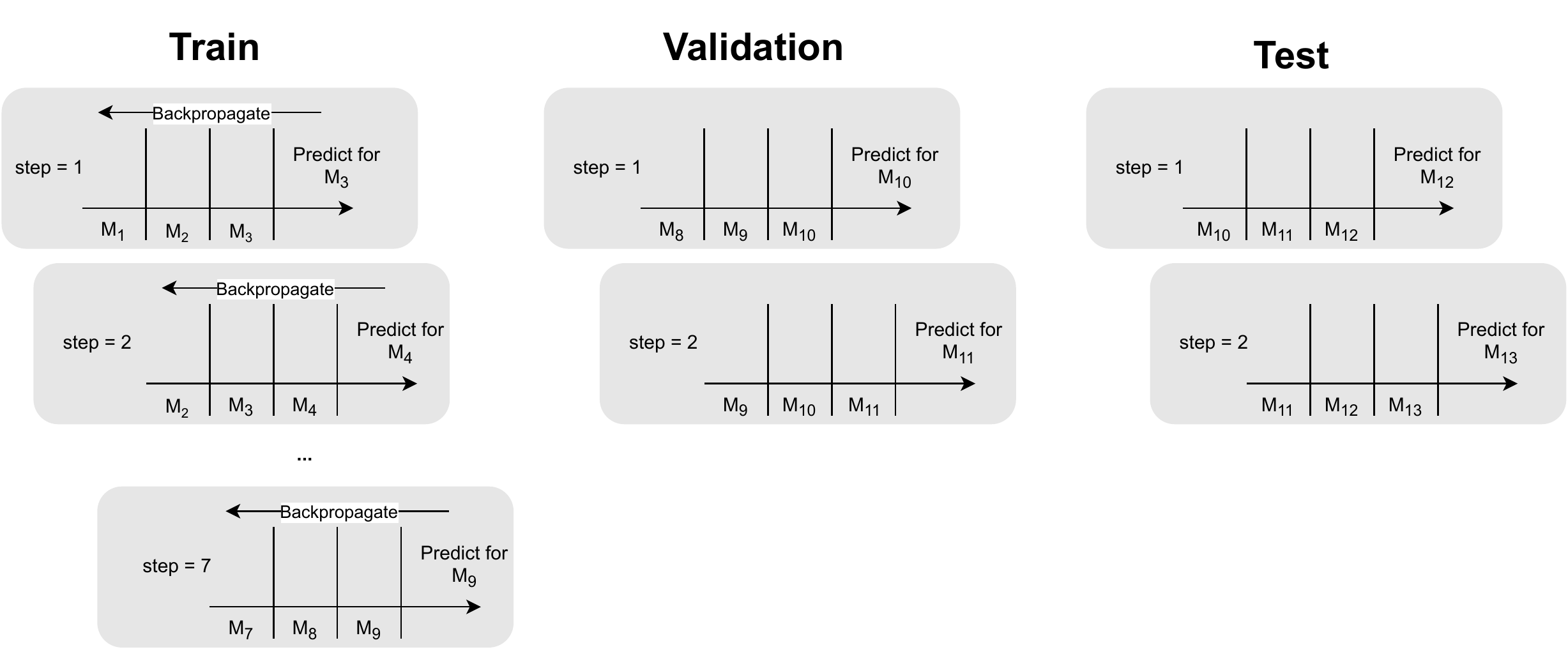}
  \caption{Train-validation-test split and model training. Windows are obtained by 1-month shift. Backpropagation happens at the end of the time window, and the prediction is made for the last month of the window.}
  \label{fig:Model_implementation}
\end{figure}

\begin{figure}[H]
\centering
  \centering
  \includegraphics[width=0.7\linewidth]{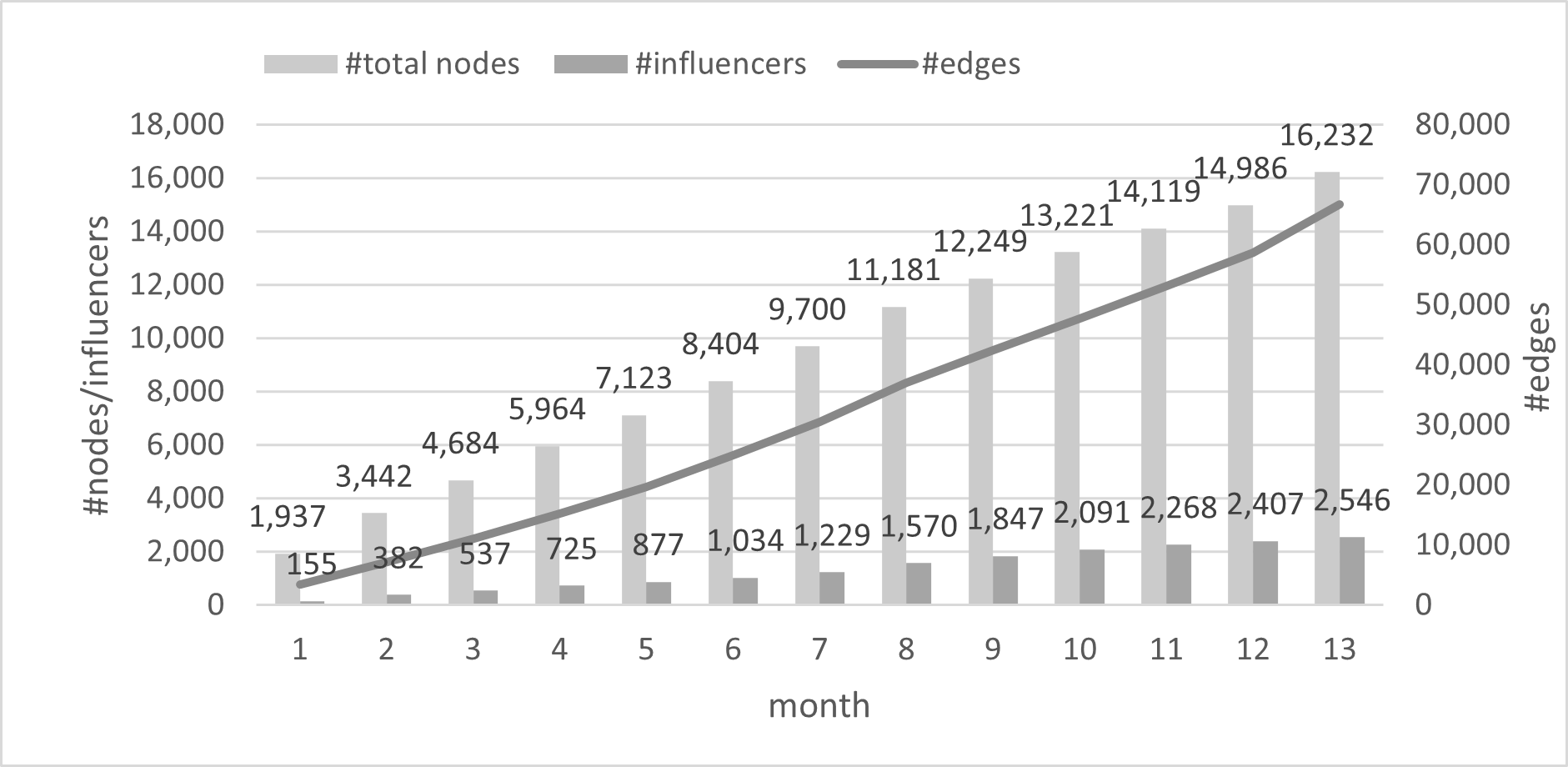}
  \caption{Network growth characteristics}
  \label{fig:Network_characteristics}
\end{figure}

\begin{table}[H]
\fontsize{8}{10}\selectfont 
  \caption{Edge types}
  \label{Edge_types}
  \centering
  \begin{tabular}{p{1.5cm}p{8cm}p{2.5cm}}
    \toprule                 \\
    Type     & Description & Data source     \\
    \midrule
    Credit card  & There exist an edge between users who used the same credit card in the app. & Delivery app   \\
    References     & There exist an edge between users who referenced/got referenced. This type of edge is created in the network snapshot of the month following the month of referral. & Credit card usage   \\
    Geohash     & There exist an edge between users who ordered more than 4 times in the close geographical proximity (Geohash 7 - 152.9m x 152.4m). & Delivery app    \\
    Contacts     & There exist an edge between users if at least one of them have the other one in the phone contacts. & Delivery app    \\
    \bottomrule
  \end{tabular}
\end{table}

\begin{table}[H]
\fontsize{8}{10}\selectfont 
  \caption{Best hyperparameter settings found by grid search: the best model has a maximal average AUC on seen and unseen nodes. Static hyperparameters: epochs = 500 (early stop = 50), learning rate = 0.0001, optimizer = ADAM \citep{adam}}
  \label{Hyperparameters}
  \centering
  \begin{tabular}{p{1.8cm}p{1.3cm}p{1.3cm}p{1cm}p{0.8cm}p{0.8cm}p{0.8cm}p{0.8cm}p{0.8cm}}
    \toprule                 \\
    Model     & Val. AUC seen nodes  & Val. AUC unseen nodes & \#hidden dimensions GNN/RNN & \#emb. GNN & \#layers GNN & SMOTE sample rate & \#heads GAT & dropout rate GNN \\
    \midrule
    GCN+LSTM & 0.774$\pm$0.004 & 0.825$\pm$0.015 & 200 & 200 & 1 & 0 & \textbackslash & 0.5  \\
    GCN+GRU & 0.862$\pm$0.004 & 0.769$\pm$0.016 & 200 & 200 & 1 & 0.75 & \textbackslash & 0.5  \\
   GAT+LSTM & 0.860$\pm$0.003 & 0.871$\pm$0.013 & 100 & 200 & 4 & 0 & 4 & 0.5   \\
    GAT+GRU & 0.880$\pm$0.003 & 0.860$\pm$0.013 & 100 & 200 & 4 & 0 & 4 & 0.5  \\
   PageRank+LSTM & 0.676$\pm$0.002 & 0.725$\pm$0.017 & 100 & \textbackslash & \textbackslash & 0 & \textbackslash & \textbackslash  \\
   PageRank+GRU & 0.818$\pm$0.003 & 0.726$\pm$0.012 & 200 & \textbackslash & \textbackslash & 0 & \textbackslash & \textbackslash \\
    Features+LSTM & 0.678$\pm$0.003 & 0.725$\pm$0.018 & 100 & \textbackslash & \textbackslash & 0 & \textbackslash & \textbackslash  \\
   Features+GRU & 0.815$\pm$0.004 & 0.721$\pm$0.018 & 200 & \textbackslash & \textbackslash & 0 & \textbackslash & \textbackslash  \\
    Static GAT & 0.650$\pm$0.004 & 0.720$\pm$0.011 & \textbackslash & 200 & 4 & 0 & 4 & 0.5  \\
    Static GCN & 0.633$\pm$0.005 & 0.665$\pm$0.016 & \textbackslash & 200 & 1 & 0 & \textbackslash & 0.5   \\
    \bottomrule
  \end{tabular}
\end{table}

\begin{table}[H]
\fontsize{8}{10}\selectfont 
  \caption{Resource types}
  \label{resources}
  \centering
  \begin{tabular}{p{4cm}p{6cm}}
    \toprule                 \\
    Resource     & Specification \\
    \midrule
    Memory per CPU &  8G \\
    CPU cores per task    &  2 \\
    Processor   &   Intel E5-2683 v4 Broadwell @ 2.1GHz \\

    \bottomrule
  \end{tabular}
\end{table}

\end{document}